\documentclass{aa}
\usepackage{natbib}
\bibpunct{(}{)}{;}{a}{}{,}
\usepackage{graphicx}
\usepackage{txfonts}
\begin{document}
\title{Heavy elements abundances in turn-off stars and early subgiants 
  in NGC~6752\thanks{Based on observations made with the ESO VLT-UT2 
    at the Paranal Observatory, Chile (ESO-LP 165.L-0263).}
}

\author{G. James\inst{1} \and P. Fran\c{c}ois\inst{1}
  \and P. Bonifacio\inst{2} \and A. Bragaglia\inst{3} \and E. Carretta\inst{4}
  \and M. Centuri\'{o}n\inst{2} \and G. Clementini\inst{3} 
  \and S. Desidera\inst{5,6} \and R. G. Gratton\inst{4} \and F. Grundahl\inst{7} 
  \and S. Lucatello\inst{4,6} \and P. Molaro\inst{2} \and L. Pasquini\inst{8} 
  \and C. Sneden\inst{9} \and F. Spite\inst{1}
}

\offprints{G. James,\\\email{Gael.James@obspm.fr}}

\institute{Observatoire de Paris - GEPI, 61 Avenue de l'Observatoire, 75014 Paris, 
  France
  \and Osservatorio Astronomico di Trieste, Via G. B. Tiepolo 11, 34131 
  Trieste, Italy
  \and Osservatorio Astronomico di Bologna, Via Ranzani 1, 40127 Bologna, Italy
  \and Osservatorio Astronomico di Padova, Vicolo dell'Osservatorio 5, 
  35122 Padova, Italy
  \and Dipartimento di Fisica, Universit\`{a} di Pisa, Piazza Torricelli 2, 
  56100 Pisa, Italy
  \and Dipartimento di Astronomia, Universit\`{a} di Padova, Vicolo 
  dell'Osservatorio 2, 35122 Padova, Italy
  \and Department of Astronomy, University of Aarhus, Bygning 520, 
  8000 Aarhus C, Denmark
  \and European Southern Observatory, Karl-Schwarzschild-Str. 2, 
  85748 Garching bei M\"{u}nchen, Germany
  \and Department of Astronomy, University of Texas at Austin, RLM 15.308, 
  C-1400, Austin, TX, USA
}

\titlerunning{Heavy elements abundances in NGC~6752}

\date{Received ??; accepted ??}

\abstract{
 High resolution spectra ($R \ge 40\,000$) for 9 main sequence turn-off stars 
 and 9 subgiants in the globular cluster NGC~6752 were acquired with UVES on the VLT-Kueyen (UT2). These 
 data have been used to determine the abundances of some heavy elements (strontium, yttrium, barium 
and europium). This paper presents for the first time accurate results for heavy elements in 
this globular cluster. We did not find any systematic effect between the abundances found in turn-off stars, 
subgiants, and giants. 
We obtain the following mean abundances for these elements in our sample (turn-off stars and subgiants): 
$\mathrm{[Sr/Fe]} = 0.06 \pm 0.16$, $\mathrm{[Y/Fe]} = -0.01 \pm 0.12$, 
$\mathrm{[Ba/Fe]} = 0.18 \pm 0.11$, and $\mathrm{[Eu/Fe]} = 0.41 \pm 0.09$. 
The dispersion in the abundance ratios of the different elements is low and can be totally explained by 
uncertainties in their derivation. 
These ratios are in agreement with the values found in field halo stars with the same metallicity. 
We did not observe any correlation between the [n-capture/Fe] ratios and the star-to-star variations 
of the O and Na abundances. The [Ba/Eu] and [Sr/Ba] ratios show clearly that this globular cluster has also 
been uniformly enriched by r- and s-process synthesis.

}

\maketitle
\section{Introduction}

Spectroscopic observations of globular cluster stars can provide many
clues to understand the formation and evolution of the early Galaxy.
As they are among the oldest stellar objects known, they can be used as
natural laboratories for testing the theory of stellar evolution. 

In a given Galactic globular cluster, stars are commonly accepted to have 
all about the same age and at least the same metallicity.
But if there doesn't seem to be evidence yet for any kind of age variation 
among stars in a single cluster, small scatters in their individual 
abundances have already been found, thus leading to the question of the 
primordial or evolutionary origin of these scatters.
Most globular clusters (GCs) seem to have a rather homogeneous iron 
abundance, 
but there is evidence of a few clusters with star-to-star variations 
in the abundances of iron-peak elements (e.g. $\omega$~Cen). 
Further striking star-to-star variations are also seen in the abundance patterns of other 
elements (e.g. CH and CN bands, Al, Na, and Mg variations, and Al-O or O-Na 
anticorrelations). The detection of heavier elements 
and the study of their abundance patterns can be used as a tool to understand 
the origin of the chemical enrichment of a globular cluster. 
Especially, a comparison of the abundances of neutron-capture elements 
(roughly $A>60$) can be very helpful to distinguish 
which chemical enrichment scenario has more probably taken place among the 
cluster stars, depending on the relative fraction of elements produced 
by the r- or the s-process. 
As almost every globular cluster chemical analysis up to now has only 
been done on giant stars, it is therefore also significant to investigate 
the possible inhomogeneity of the abundances in the different evolutionary 
sequences of a globular cluster.

This work is part of the ESO-Large Program 165.L-0263 (PI: R. G. Gratton), 
which is dedicated to the analysis of high-resolution spectra 
($R \ge 40\,000$) of a large sample of dwarfs near the turn-off and 
early subgiants at the base of the red giant branch (RGB) in at least three 
globular clusters (47~Tuc, NGC~6397 and NGC~6752). The first results
of this program \citep{Gra2001} already showed clearly 
-and for the first
time- the O-Na anticorrelation among turn-off stars and early subgiants in
NGC~6752, ruling out the hypothesis of mixing as being the origin of these 
anomalies. A more recent work has been done by \citet{Boni2002} on the 
lithium content of NGC~6397, its 
possible primordial origin in this cluster, and its influence on the standard 
big-bang model. 

Previous precise abundance determinations in globular clusters have only 
been done for bright giant stars ($V \sim 11$--12), and the few attempts 
to derive abundances in fainter or less evolved stars \citep[e.g.][]{Boes1998,King1998} 
obtained only very low $S/N$ spectra, and thus not really reliable measurements of the 
n-capture elements. UVES, the echelle-spectrograph at the VLT-UT2 
(Kueyen), is one of the few instruments able to obtain spectra with higher $S/N$ values 
for such faint stars ($V \sim 16$--17), which allows to measure accurate abundances 
also for heavy elements.

Concerning the cluster NGC~6752, there are only a few 
references about abundance determinations, most of them being 
metallicity (in this case, iron abundances) determinations \citep{Zinn1984,Min1993,Carr1997}, 
and some others 
beeing abundance analyses including several other elements \citep{Norris1995a,Grun2002,Yong2003}. 
All these studies 
have been done on bright giant stars, and the only few 
results on heavy elements are reported in the paper of \citet{Norris1995a}.

In this paper, we present for the first time accurate results for abundances 
of several heavy elements (Sr, Y, Ba, and Eu) in turn-off stars and 
early subgiants (at the base of the RGB) in a globular cluster, NGC~6752.

\section{Observations}

\begin{figure}
  \centering
  \includegraphics[width=8cm]{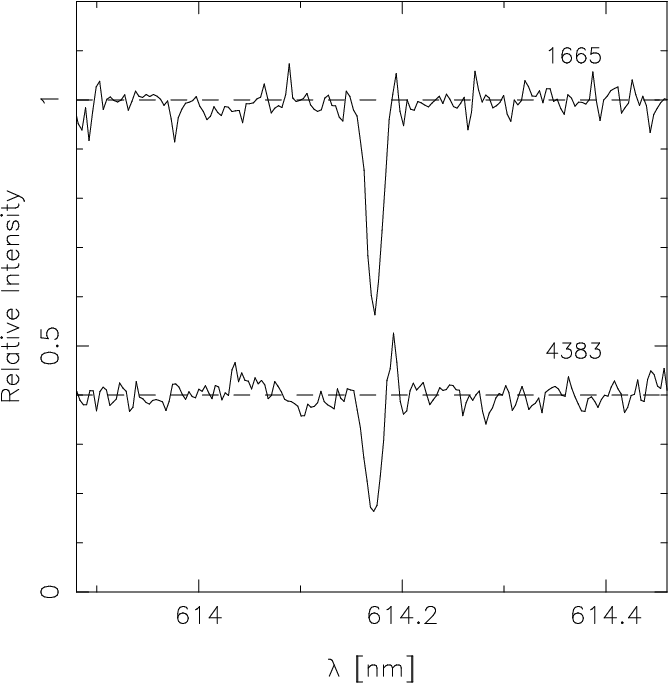} \\
  \vspace{0.5cm}
  \includegraphics[width=8cm]{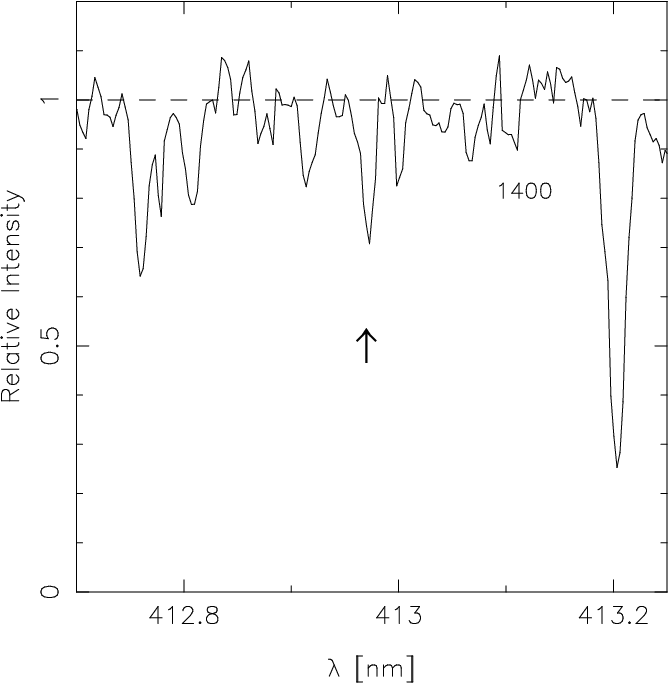}
  \caption{\textit{Top}: UVES (VLT-UT2) normalized spectra of NGC~6752 stars $\#$1665 
    (subgiant) and $\#$4383 (dwarf), centered on the \ion{Ba}{ii} (614.17 nm) 
    line, showing a direct comparison of the strength of the lines between 
    subgiants and dwarfs in the cluster. The spectrum of star $\#$4383 has been 
    offsetted to avoid overlap. 
    \textit{Bottom}: Spectral region of star $\#$1400 (subgiant), 
    centered on the \ion{Eu}{ii} (412.97 nm) line.
  }
  \label{FigBaEu_raw}
\end{figure}

The observations were carried out in several runs at the Kueyen (VLT-UT2) 
Telescope with the UVES high-resolution echelle spectrograph in June and 
September 2000, and they have already been described by \citet{Gra2001}. 
For the cluster NGC~6752, data have been obtained for 9 
dwarfs around the turn-off (TO) point, and 9 subgiants at the base of the RGB.

The observed stars are listed in Table~\ref{TableData}. Star identifications 
and photometry are from \citet{Grun2000}. Exposure times range 
from about 1 hour for the brightest subgiants to about 4 hours (split in 3 to 
4 exposures) for the faintest turn-off stars. The observations were done 
with the UVES Dichroic \#2 mode, which allows to cover a wide spectral 
range (350--470 nm for the blue spectra, and 570--870 nm for 
the red ones). The resolution ($R \equiv \lambda / \Delta \lambda$) is 
always over 40\,000, depending mainly on the 
slit width \citep[see][]{Gra2001}. 

In Table~\ref{TableData}, we give $S/N$ ratios per pixel around 450 and 650 nm 
(there are $\sim$5 pixels per resolution element), 
near the observed lines of strontium, yttrium, barium, and europium 
(\ion{Sr}{ii}: 407.77 and 421.55 nm; 
\ion{Y}{ii}: 395.03 and 439.80 nm; 
\ion{Ba}{ii}: 455.40, 585.37, 614.17 and 649.69 nm; 
\ion{Eu}{ii}: 412.97 and 420.50 nm). $S/N$ ratios range from $\sim$15 in 
the worst 
cases of the blue spectra (for the turn-off star \#4661), to $\sim$65 in 
the red spectra (subgiant \#1400, and turn-off star \#4907).
Figure~\ref{FigBaEu_raw} shows typical examples of our spectra in 
the regions of the \ion{Ba}{ii} line at 614.17 nm, and the \ion{Eu}{ii} 
line at 412.97 nm.

\section{Abundance analysis}

\begin{table*}
  \centering
  \caption[]{Data for our NGC~6752 Stars. Star identifications and photometry 
    are from \citet{Grun2000}.}
  \label{TableData}
  \begin{tabular}{lccccrrrrr}
    \hline\hline
    \noalign{\smallskip}
    \hbox{Star  \#}  & $V$ & $b-y$ & $S/N$   & $S/N$   & [Fe/H] & [Sr/Fe] & [Y/Fe] & [Ba/Fe] & [Eu/Fe]\\
    &     &       & @450 nm & @650 nm &        &         &        &         &        \\
    \noalign{\smallskip}
    \hline
    \noalign{\smallskip}
    \multicolumn{10}{c}{Subgiants}\\
    \multicolumn{10}{c}{($T_\mathrm{eff} = 5347\ \mathrm{K}$; $\log g = 3.54$; $\mathrm{[Fe/H]} = -1.50$; $\xi = 1.1\ \mbox{km s}^{-1}$)} \\
    \noalign{\smallskip}
    \hline
    \noalign{\smallskip}
    1400       & 15.89   & 0.50  &   49     &   65     &  --1.40  & --0.12    &   0.00   &   0.10    &  0.37    \\
    1406       & 15.90   & 0.48  &   54     &   58     &  --1.47  &   0.03    & --0.02   &   0.20    &  0.42    \\
    1445       & 15.89   & 0.50  &   36     &   57     &  --1.56  &   0.00    & --0.09   &   0.33    &  0.51    \\
    1460       & 15.94   & 0.47  &   35     &   58     &  --1.42  & --0.04    & --0.23   &   0.20    &  0.31    \\
    1461       & 15.95   & 0.49  &   26     &   44     &  --1.40  & --0.07    & --0.10   &   0.23    &  0.25    \\
    1481       & 15.95   & 0.46  &   42     &   62     &  --1.52  &   0.13    &   0.00   &   0.30    &  0.46    \\
    1563       & 16.03   & 0.47  &   33     &   52     &  --1.60  &   0.06    &   0.22   &   0.36    &  0.54    \\
    1665       & 16.04   & 0.48  &   30     &   47     &  --1.50  &   0.07    &   0.03   &   0.27    &  0.36    \\
    202063     & 15.94   & 0.47  &   27     &   50     &  --1.55  & --0.11    &   0.13   &   0.23    &  0.40    \\
    \noalign{\smallskip}
    \hline
    \noalign{\smallskip}
    $< \mathrm{mean} >$ &   &    &          &          & --1.49   & --0.01    & --0.01   &   0.25    &  0.40    \\
    $\sigma$ (std. dev.)   &  &  &          &          &   0.07   &   0.09    &   0.13   &   0.08    &  0.09    \\
    \noalign{\smallskip}
    \hline
    \noalign{\smallskip}
    \multicolumn{10}{c}{Dwarfs} \\
    \multicolumn{10}{c}{($T_\mathrm{eff} = 6226\ \mathrm{K}$; $\log g = 4.28$; $\mathrm{[Fe/H]} = -1.50$; $\xi = 0.7\ \mbox{km s}^{-1}$)}         \\
    \noalign{\smallskip}
    \hline
    \noalign{\smallskip}
    4341       & 17.15   & 0.35  &   20     &   44     &  --1.48  & --0.07    & --0.13   &   0.05    & $<$0.52 \\
    4383       & 17.11   & 0.36  &   39     &   59     &  --1.45  &   0.11    &   0.00   &   0.14    & $<$0.44 \\
    4428       & 17.14   & 0.37  &   42     &   61     &  --1.59  &   0.30    & $<$0.00  &   0.20    & $<$0.63 \\
    4458       & 17.16   & 0.34  &   24     &   45     &  --1.50  & --0.19    & $<$0.03  &   0.15    & $<$0.62 \\
    4661       & 17.22   & 0.34  &   16     &   26     &  --1.42  &   0.28    & $<$0.12  & --0.01    & $<$0.67 \\
    4907       & 17.20   & 0.35  &   52     &   66     &  --1.52  &   0.29    &   0.09   &   0.18    &    0.53 \\
    5048       & 17.28   & 0.35  &   36     &   62     &  --1.47  &   0.27    &   0.07   &   0.14    &    0.41 \\
    200613     & 17.20   & 0.38  &   19     &   34     &  --1.35  &   0.19    & --0.19   & --0.03    & $<$0.44 \\
    202316     & 17.28   & 0.35  &   35     &   51     &  --1.53  & --0.06    &   0.01   &   0.21    & $<$0.47 \\
    \noalign{\smallskip}
    \hline
    \noalign{\smallskip}
    $< \mathrm{mean} >$ & &      &          &          & --1.48   &   0.12    & --0.03   &   0.11    &  0.47    \\
    $\sigma$ (std. dev.)  &   &  &          &          &   0.07   &   0.19    &   0.11   &   0.09    &  0.08    \\
    \noalign{\smallskip}
    \hline
  \end{tabular}
\end{table*}

\subsection{Model atmospheres and stellar parameters}

The adopted model atmospheres (OSMARCS, LTE) have been computed using the grid 
defined by \citet{Edv1993}, with an updated version 
of the MARCS code of \citet{Gus1975} with improved 
UV-line blanketing \citep[see also][]{Edv1994}. In 
these computations, we used the solar abundances of \citet{Grev2000}.

Our sample stars near the turn-off are all very close to each other in the 
Str\"{o}mgren color-magnitude diagram of NGC~6752, and so are also the sample 
stars at the base of the RGB. All the stars in each one of these 
two sets of data have consequently very similar stellar parameters: 
effective temperatures ($T_\mathrm{eff}$), surface gravities ($\log g$), 
and microturbulent velocities ($\xi$). 
\citet{Gra2001} already showed clearly that we could use 
two sets of mean atmospheric parameters: one for the TO stars, and another 
one for the subgiants. In the present work, we used these mean parameters, 
and adapted them to fit the data with our models.

To check the parameters given by \citet{Gra2001}, we 
first used two spectra: the average of the 9 TO stars, and the average of 
the 9 subgiants. The $S/N$ ratios (per pixel) of these summed spectra are much 
higher than those of the single spectra ($\sim$60 and $\sim$80 for the blue 
TO and subgiants average spectra around 450 nm; $\sim$100 and $\sim$140 
for the red TO and subgiants average spectra near 650 nm; see 
Table~\ref{TableData} for the individual $S/N$ ratios), which allows an 
accurate equivalent width measurement of many reliable 
\ion{Fe}{i} and \ion{Fe}{ii} lines.
 
\citet{Gra2001} obtained their effective temperatures 
using Str\"{o}mgren photometry from \citet{Grun2000}, and by fitting the wings 
of the H$_\alpha$ profiles of the spectra. They compared the resulting 
effective temperatures to a calibration based on the work of \citet{Al1996} for MS-stars. 
In this work, the effective temperature for our two resulting spectra was checked by 
assuming that there is no trend between the \ion{Fe}{i} equivalent widths
and the excitation potentials, and by fitting the wings of the H$_\alpha$ 
profile of the spectra. We also checked the gravities and microturbulent 
velocities by comparing theoretical curves of growth with observational 
curves of growth. The gravities were verified with the 
ionization equilibrium of \ion{Fe}{i} and \ion{Fe}{ii}, and the microturbulent velocities were 
checked imposing the \ion{Fe}{i} abundances to be independent of the 
equivalent width of the lines. Finally, we found that the parameters used by 
\citet{Gra2001} were compatible with these verifications within the error bars given in 
that paper.

These sets of parameters were then used to recompute the Fe abundance\footnote{We 
adopt here the usual spectroscopic notations that 
$\mathrm{[A/B]} \equiv {\log}_{10} (N_\mathrm{A}/N_\mathrm{B}) - {\log}_{10} {(N_\mathrm{A}/N_\mathrm{B})}_{\sun}$, 
and that 
$\log \epsilon (\mathrm{A}) \equiv {\log}_{10} (N_\mathrm{A}/N_\mathrm{H}) + 12.0$, 
for elements A and B. We assume also that metallicity is 
equivalent to the stellar $\mathrm{[Fe/H]}$ value.} with our OSMARCS code, and we 
found $\mathrm{[Fe/H]} = -1.50$, which we finally used as input parameter 
in our detailed computations.

As a next step, we checked for each single star that these parameters were 
giving coherent results by computing the abundances of up to $\sim$30 lines of 
\ion{Fe}{i} and \ion{Fe}{ii} weaker than 100 m\AA. At last, we adopted the 
following two sets of parameters: 
\{$T_\mathrm{eff} = 5347\ \mathrm{K}$; $\log g = 3.54$; 
$\mathrm{[Fe/H]} = -1.50$; $\xi = 1.1\ \mbox{km s}^{-1}$\} for the 
subgiants, 
and \{$T_\mathrm{eff} = 6226\ \mathrm{K}$; $\log g = 4.28$; 
$\mathrm{[Fe/H]} = -1.50$; $\xi = 0.7\ \mbox{km s}^{-1}$\} for the 
turn-off stars (see Table~\ref{TableData}).

\subsection{Iron abundances}

For iron lines, we deduced abundances from equivalent width measurements made 
with an automatic line fitting procedure based on the algorithms of 
\citet{Char1995}, which perform both line detection and 
Gaussian fits of unblended lines. Although many more lines are visible on the 
spectra, we made a selection of 30--40 (depending on the quality of the 
spectra) unblended \ion{Fe}{i} lines with 
equivalent widths lower than 100 m\AA, and we kept all the detected unblended 
\ion{Fe}{ii} lines. The results for individual stars are listed in 
Table~\ref{TableData}. The iron abundances are the average of the \ion{Fe}{i} and 
\ion{Fe}{ii} abundances.

Previous iron abundance determinations have already been done for this 
cluster. For bright giant stars, \citet{Zinn1984}, \citet{Min1993}, and 
\citet{Norris1995a} found respectively 
$\mathrm{[Fe/H]} = -1.54$, $\mathrm{[Fe/H]} = -1.58$ (3 stars), 
and $\mathrm{[Fe/H]} = -1.54$ (6 stars), 
while \citet{Carr1997} found a 
higher value for their 4 observed giants ($\mathrm{[Fe/H]} = -1.42$). 
\citet{Gra2001} confirm this metallicity for our sample 
($\mathrm{[Fe/H]} = -1.42$). More recently, \citet{Grun2002} and 
\citet{Yong2003} published a metallicity of $\mathrm{[Fe/H]} = -1.62$, 
respectively for 21 RGB bump stars, and 20 bright giant stars in this cluster. 
The average value of $\mathrm{[Fe/H]} = -1.49 \pm 0.07$ (standard deviation 
around the mean value, see Table \ref{TableData}) for our whole 
sample is fully compatible with all these previous values within the errors 
that could have been made on our atmospheric parameters 
(see Section~\ref{errors} hereafter).

The differences in iron abundance between the different previous analyses on this cluster 
have been discussed by \citet{KrIv2003} and seem to be due mainly to 
differences in the adopted models (MARCS, Kurucz...).  
And more precisely, concerning the slight difference (0.07 dex) of our value and the previous metallicity 
published by \citet{Gra2001} from the same data, it is partly due to our different $T(\tau)$ laws 
(temperature variation as a function of the optical depth), leading to a difference 
in the temperature at the depth of line formation, and finally to a small shift in the metallicity 
(we used OSMARCS models, while the previous analysis has been done using Kurucz models). 
Moreover, when comparing our Fe abundances with those obtained in \citet{Gra2001}, it should be noted 
that these last are the abundances from \ion{Fe}{i} alone. Those obtained from \ion{Fe}{ii} lines are 
a bit lower (--1.49 for the TO stars, and --1.57 for the stars at the base of the RGB). If the same average 
than in the present paper is done, the difference between the two determinations would be very small 
($\sim$0.02 dex; the exact value depends on the weights given when computing the average). However, it should 
also be noticed that the value used in \citet{Gra2001} is that appropriate for the purpose of obtaining the 
ages \citep{Gra2003} because also for the field stars we only consider \ion{Fe}{i}.

But this is not a critical point in this work because we are presenting abundances ratios. 
Apart from that, we confirm the small dispersion in iron 
abundance: the standard deviation\footnote{Here we call standard deviation 
the square root of the variance: $\sigma(x_1 \ldots x_N) = \sqrt{\mathrm{Var}(x_1 \ldots x_N)}$, 
where the variance is given 
by $\mathrm{Var}(x_1 \ldots x_N) = \frac{1}{N-1} \sum_{i=1}^{N}{(x_i - \bar{x})}^2$, 
and $\bar{x}$ is the mean value of $(x_1 \ldots x_N)$.} for the whole sample (TO stars and subgiants) 
is 0.07 (we obtain 0.10 from the data published in \citealt{Gra2001}).

\subsection{Heavy elements abundances}

\begin{figure}
  \centering
  \includegraphics[width=8cm]{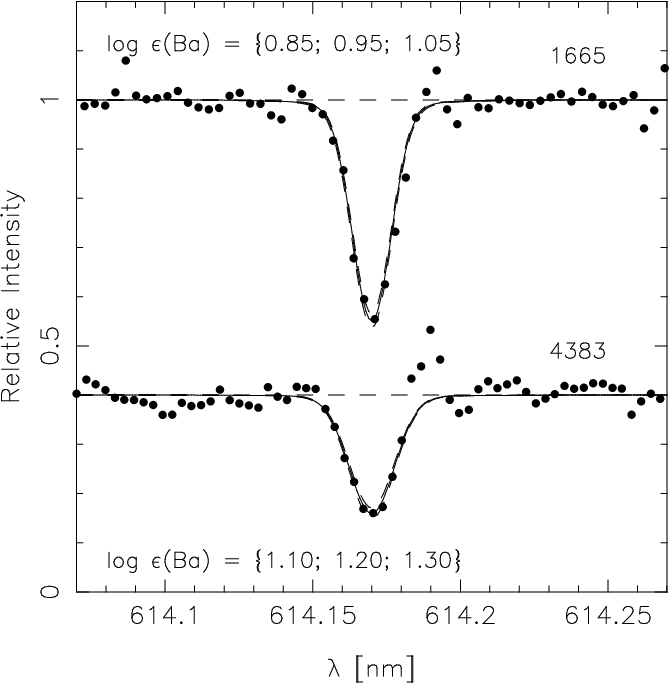} \\
  \vspace{0.5cm}
  \includegraphics[width=8cm]{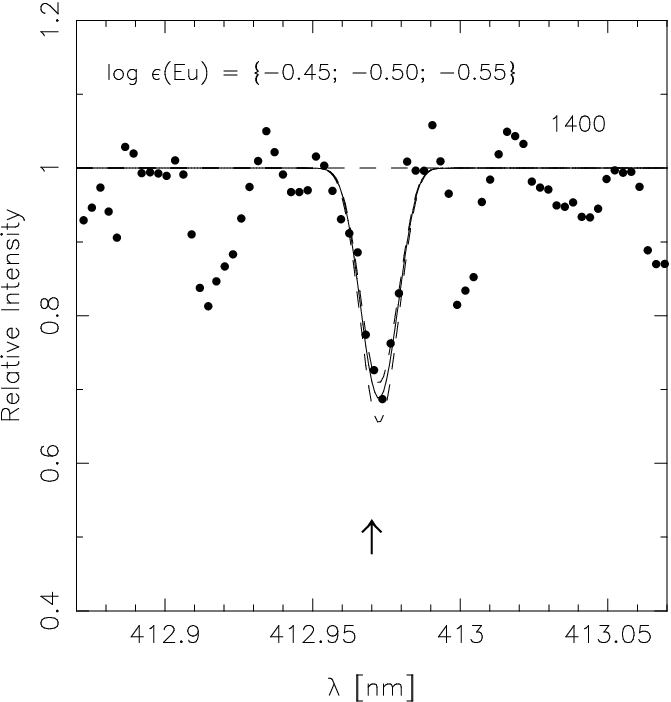}
  \caption{\textit{Top}: Synthetic spectra obtained for the 
    \ion{Ba}{ii} (614.17 nm) line of stars $\#$1665 (subgiant) and $\#$4383 
    (dwarf), with additional offset. 
    \textit{Bottom}: Synthetic spectrum of the \ion{Eu}{ii} 
    (412.97 nm) line in star $\#$1400 (subgiant). 
    \textit{Dots}: observations, \textit{lines}: synthetic spectra.
  }
  \label{FigBaEu_fit}
\end{figure}

Strontium and yttrium abundances were computed from equivalent width 
measurements in the same way as our iron abundances. The excitation 
potentials and $\log gf$ for the \ion{Sr}{ii} and \ion{Y}{ii} lines were 
taken from \citet{Sneden1996}. As barium and europium 
transitions are generally affected 
by hyperfine structure (hereafter hfs; e.g. see \citealt{McW1995}; 
\citealt{McW1998}, and references therein), we computed a 
synthetic spectrum over several {\AA} and compared it interactively to the 
observed one. We used the latest version of the synthetic spectrum codes of 
\citet{Spite1967}. Figure~\ref{FigBaEu_fit} shows typical comparisons 
between synthetic and observed spectra for the \ion{Ba}{ii} line at 614.17 nm 
and for the \ion{Eu}{ii} line at 412.97 nm for dwarfs and/or subgiants of our 
sample. The hfs parameters for the \ion{Ba}{ii} and \ion{Eu}{ii} lines have 
been taken respectively from \citet{McW1998} and \citet{Lawler2001}.

For strontium, mean abundances were obtained from at 
least one line at 407.77 nm or at 421.55 nm, and in most cases from both lines. 
For yttrium, we could almost always detect the lines at 395.03 and 
439.90 nm in the subgiants, and at least the line at 395.03 nm in the dwarfs. 
For barium, we had at least three lines at 455.40, 614.17 and 
649.69 nm, and sometimes also a third line at 585.37 nm. In the case of europium, 
we usually could detect the two expected lines at 412.97 and 420.50 nm in the 
subgiants, but for most of the TO stars, we could only give upper limits for both lines. 

The abundances of these four heavy elements are given in 
Table~\ref{TableData}. We obtain the following average values (with indication 
of the standard deviation): 
$\mathrm{[Sr/Fe]} = 0.06 \pm 0.16$, $\mathrm{[Y/Fe]} = -0.01 \pm 0.12$, 
$\mathrm{[Ba/Fe]} = 0.18 \pm 0.11$, and $\mathrm{[Eu/Fe]} = 0.41 \pm 0.09$ for 
our sample stars. Figure~\ref{FigAbond} displays results for [X/Fe] as a 
function of [Fe/H], where X is either Sr, Y, Ba, or Eu. These abundance 
patterns will be discussed in the following sections (see Section~\ref{abpatterns}). 

Our computations have been done using LTE model atmospheres. But it is clear 
that NLTE can affect heavy elements, in particular \ion{Ba}{ii} and \ion{Sr}{ii} lines, for metal-poor 
stars in the metallicity range $-1.90 \le \mathrm{[Fe/H]} \le -1.20$ \citep[e.g.][]{Mash1999}. 
Detailed computations of the NLTE effects have been done respectively by \citet{Mash1999} and 
\citet{Mash2000,Mash2001} for \ion{Ba}{ii}, \ion{Eu}{ii} and \ion{Sr}{ii} lines. For 
metallicities around -1.50, in most of the cases the NLTE corrections do not exceed 0.15 dex 
in intensity. However, the intensity of these effects depends strongly on the observed line 
and on the atmospheric parameters, and since the uncertainties on our atmospheric parameters 
(see Section~\ref{errors} and Table~\ref{TableErrors}) combined with the uncertainties on 
the EW measurements (or on the fits for the abundances deduced from synthetic spectra) 
lead to total errors of $\sim$0.15 dex, it is not certain that there would 
be a significant change in the abundances of our sample.

\subsection{Error estimations}
\label{errors}

\begin{table}
  \centering
  \caption[]{Error estimates for the sample stars.}
  \label{TableErrors}
  \begin{tabular}{lccccc}
    \hline\hline
    \noalign{\smallskip}
    El.  & $N^\mathrm{a}$ & $\Delta T_\mathrm{eff}$ & $\Delta \log g$ & $\Delta \xi$     & $\Delta (\mathrm{Tot})^\mathrm{b}$\\
    &                & +100 K                  & +0.2 dex        & +0.2 km s$^{-1}$ &    \\
    \noalign{\smallskip}
    \hline
    \noalign{\smallskip}
    \multicolumn{6}{c}{Subgiants (e.g. \#1400)}\\
    \noalign{\smallskip}
    \hline
    \noalign{\smallskip}
    \ion{Fe}{i}  & 33 &  +0.09 & --0.02 & --0.05 &   0.10 \\
    \ion{Fe}{ii} & 11 & --0.01 &  +0.07 & --0.05 &   0.09 \\
    \noalign{\smallskip}
    \hline
    \noalign{\smallskip}
    \ion{Sr}{ii} &  1 &  +0.04 & --0.02 &  +0.02 &  0.05 \\
    \ion{Y}{ii}  &  2 & --0.01 &  +0.05 & --0.04 &  0.06 \\
    \ion{Ba}{ii} &  3 &  +0.04 &  +0.01 & --0.05 &  0.06 \\
    \ion{Eu}{ii} &  2 &   0.00 &  +0.05 & --0.04 &  0.06 \\
    \noalign{\smallskip}
    \hline
    \noalign{\smallskip}
    \multicolumn{6}{c}{Dwarfs (e.g. \#4907)} \\
    \noalign{\smallskip}
    \hline
    \noalign{\smallskip}
    \ion{Fe}{i}  & 27 &  +0.09 & --0.04 & --0.05 &  0.11 \\
    \ion{Fe}{ii} &  8 &  +0.01 &  +0.06 & --0.03 &  0.07 \\
    \noalign{\smallskip}
    \hline
    \noalign{\smallskip}
    \ion{Sr}{ii} &  1 &  +0.05 & --0.04 &  +0.01 &  0.06 \\
    \ion{Y}{ii}  &  2 &   0.00 &  +0.06 &  +0.02 &  0.06 \\
    \ion{Ba}{ii} &  3 &  +0.04 & --0.02 & --0.06 &  0.07 \\
    \ion{Eu}{ii} &  2 &   0.00 &  +0.06 &  +0.03 &  0.07 \\
    \noalign{\smallskip}
    \hline
  \end{tabular}
  \begin{list}{}{}
  \item[$^\mathrm{a}$] For \ion{Fe}{i}, we only took into account lines 
    with $W_\lambda \le 100 \, \mathrm{m\AA}$.
  \item[$^\mathrm{b}$] $\Delta (\mathrm{Tot}) = \Delta (T_\mathrm{eff} \mbox{, } \log g \mbox{, } \xi)$ is the quadratic sum 
    $\sqrt{ \sum \Delta ^2   }$ of all the individual uncertainties linked to the atmospheric parameters.
  \end{list}
\end{table}

For the iron abundances, we assumed that the total error budget was due to 
random uncertainties in the measurement of the equivalent widths, and the 
errors made on the stellar parameters. When $N \ge 2$ lines of a given element 
(here \ion{Fe}{i} or \ion{Fe}{ii}) are observed, the random uncertainties can be 
computed as the standard deviation around the mean abundance. 
The errors linked to the uncertainties on the stellar atmosphere 
parameters were estimated assuming the following variations: 
$\Delta T_\mathrm{eff} = \pm 100\ \mathrm{K}$, 
$\Delta \log g = \pm 0.2\ \mathrm{dex}$, and 
$\Delta \xi = \pm 0.2\ \mbox{km s}^{-1}$.

For the heavy elements, as there are only a few lines detected per element, 
we only computed the errors for [X/Fe] due to the uncertainties 
in our choice of stellar parameters. The total error budget on the stellar 
atmosphere parameters $\Delta (T_\mathrm{eff} \mbox{, } \log g \mbox{, } \xi)$ 
is then the quadratic sum of the errors on the individual parameters.

Table~\ref{TableErrors} summarizes the errors estimates for one turn-off star 
and for one subgiant in NGC~6752. Computations of the errors for the other 
stars of our sample give similar results. The total error given in the last column of 
Table~\ref{TableErrors}, slightly smaller than the scatter given in 
Table~\ref{TableData}, does not take into acount the uncertainty coming from 
the measurements of the equivalent widths.

\section{Discussion}

\subsection{Abundance patterns}
\label{abpatterns}

\begin{figure*}
  \centering
  \includegraphics[width=17cm]{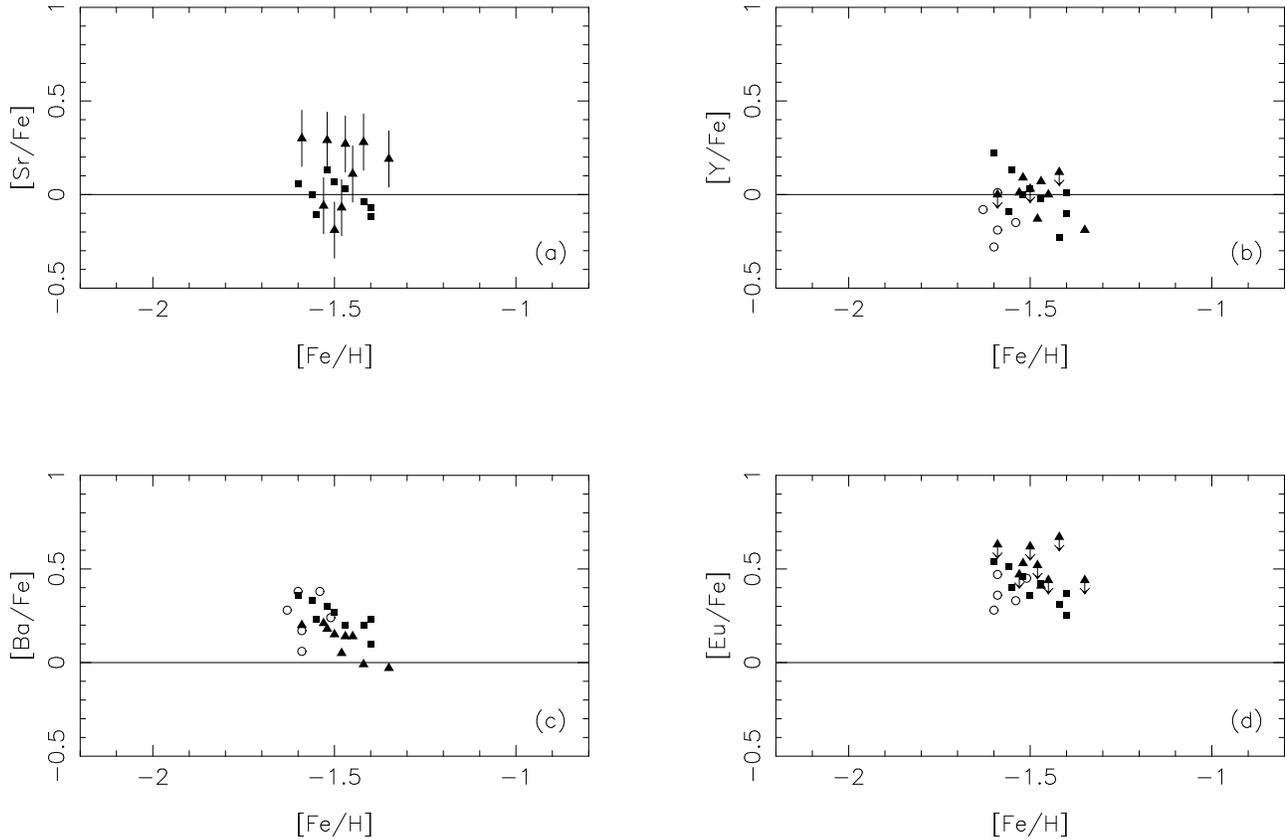}
  \caption{$\mathrm{[X/Fe]}$ ratios for the heavy elements in NGC~6752. 
    \textit{Squares:} stars at the base of the RGB. 
    \textit{Triangles:} dwarfs near the turn-off point.
    \textit{Arrows:} upper limits for the abundances in the dwarfs.
    \textit{Circles:} data for giants from \citet{Norris1995b} 
    and \citet{Norris1995a}.
  }
  \label{FigAbond}
\end{figure*}

Figure~\ref{FigAbond} displays the abundances ratios [Sr/Fe], [Y/Fe], [Ba/Fe] and [Eu/Fe] as a 
function of [Fe/H] for our sample stars.  We have added the results from \citet{Norris1995a} 
on these figures. In order to get an homogeneous sample of data, we have 
recomputed their abundances in the following way. We took their published list of equivalent widths 
for \ion{Fe}{i}, \ion{Fe}{ii},\ion{Y}{ii}, \ion{Ba}{ii} and \ion{Eu}{ii} 
\citep{Norris1995b} and computed new abundances using our own lines parameters 
($\log gf$)  and the stellar atmosphere model (OSMARCS) used for our computations. 
The data of \citet{Norris1995a} are displayed on the figures with open circles. 
In the four panels, the dwarf stars are represented by triangles whereas the subgiants are represented 
by squares. An inspection of these figures does not seem to show large systematic differences between 
abundances obtained from  dwarfs and subgiants.

\begin{itemize}
\item{Figure~\ref{FigAbond}a: Strontium}

For strontium, we have a significant spread in the abundance ratios compared to the values found for 
the other elements. It is interesting to note that this effect is also found in halo field stars 
(e.g. see \citealt{Ryan1991}; \citealt{Burris2000}, and references therein). 
For dwarf stars, we have a larger spread than for subgiants (see also Table \ref{TableData}). 
This effect can be totally explained by the fact that the spectra for dwarfs are of lower quality. 
Although the \ion{Sr}{ii} lines are rather strong, the error on the determination 
of the equivalent widths comes mostly from the difficulty to estimate the continuum level. 
In the plot, we have added error bars for the dwarfs taking into account this additional error 
($\sim 10$--15\%). We obtain a solar value for the mean [Sr/Fe] ratio of $0.06 \pm 0.16$. 

\item{ Figure~\ref{FigAbond}b: Yttrium}

We find [Y/Fe] ranging from -0.23 to 0.22. There is no indication of an effect 
as a function of the spectral class of the star. We obtain a solar value of $-0.01 \pm 0.12$ for 
the mean [Y/Fe] ratio. This ratio is totally compatible with the values found in halo stars at 
intermediate metallicity \citep{Burris2000,Fulb2000}.

\item{Figure~\ref{FigAbond}c: Barium}

We found an overabundance of the [Ba/Fe] ratio of +0.18 dex with a small offset (0.14 dex) between 
dwarfs and subgiants. The origin of this difference remains unclear. It could partly come from the 
difference in the quality of the spectra. In any case, it is clear that our sample is not large enough 
to draw final conclusions on this matter. 
This result is not very different from what has been found before for this 
cluster (see \citealt{Norris1995a}, who found an average value of 
$\mathrm{[Ba/Fe]} = 0.25$), and it is also compatible with previous analyses of metal-poor stars 
in the same metallicity range \citep{Burris2000,Fulb2000,Mash2003}. 
It is interesting to compare the present results for our almost unevolved sample of stars to 
the abundance ratios found in giants in other globular clusters. \citet{Sneden1997}, 
\citet{Sneden2000}, and \citet{Armos1994} have found mean [Ba/Fe] 
ratios ranging from --0.29 to 0.12 in globular clusters with metallicities between --1.17 and --2.41. 
Our mean value $\mathrm{[Ba/Fe]} = 0.18 \pm 0.11$ is in fair agreement with these values. 

\item{Figure~\ref{FigAbond}d: Europium}

Europium can hardly be detected in the TO stars of NGC~6752, but the few reliable detections are 
compatible with the abundances found in the subgiants of the cluster (see also Table~\ref{TableData}). 
There is a clear overabundance of the ratio [Eu/Fe] with a mean value of $0.41 \pm 0.09$ dex. 
This value is in agreement with the results of \citet{Shet1996} for M~71, M~5, M~13, and M~92, 
which are all overabundant in Eu by about the same value, and also with the ratios found in other 
metal-poor stars at intermediate metallicity \citep{Burris2000,Fulb2000,Mash2003}. 
It is interesting to note that the dispersion of the [Eu/Fe] ratio among 
the four globular clusters studied by \citet{Shet1996} is much smaller than the dispersion of 
the [Ba/Fe] ratio. In the present case, it is not possible to reach any conclusion concerning 
the different dispersion in [Ba/Fe] and [Eu/Fe] from our data (see Table~\ref{TableData}).
\end{itemize}

To summarize, we did not find any large systematic difference in the abundance ratios for subgiants and 
turn-off stars. Our values are compatible with the abundance ratios found by \citet{Norris1995a}, 
and also with the values given by several analyses of metal-poor stars at intermediate metallicity. 
For the Ba and Eu abundances, the slight difference between our analysis and the paper of 
\citet{Norris1995a} may be due to the fact that they did not consider the hyperfine structure 
of the transitions for these elements. No obvious correlation was detected in the comparison of the 
[n-capture/Fe] abundance ratios as a function of [O/Fe], [Mg/Fe], and [Na/Fe] 
(taken from \citealt{Gra2001}).

\subsection{Neutron-capture elements ratios}

\begin{figure}
  \centering
  \includegraphics[width=8cm]{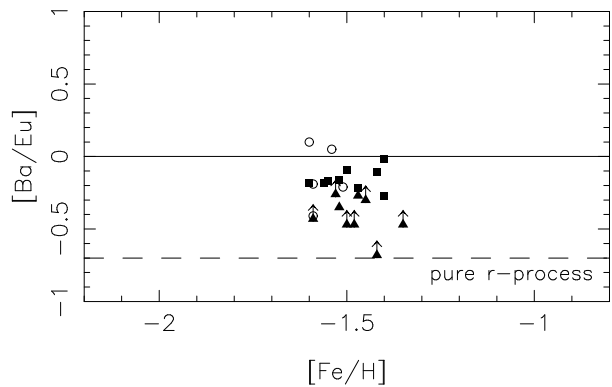}
  \caption{[Ba/Eu] vs. [Fe/H]. The symbols are the same than in 
    Figure~\ref{FigAbond}.
  }
  \label{FigBaEu}
\end{figure}

\begin{figure}
  \centering
  \includegraphics[width=8cm]{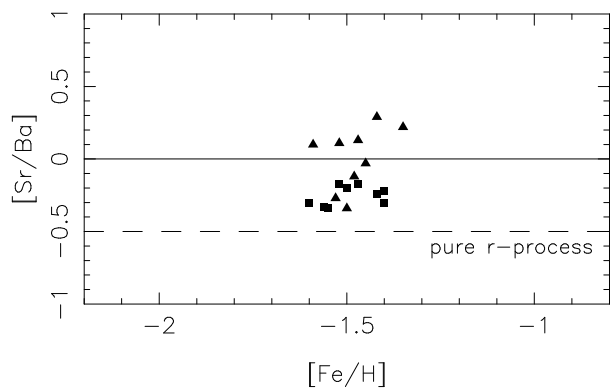}
  \caption{[Sr/Ba] vs. [Fe/H]. Symbols are the same than in 
    Figure~\ref{FigAbond}.}
  \label{FigSrBa}
\end{figure}

Figure~\ref{FigBaEu} shows [Ba/Eu] as a function of [Fe/H]. The horizontal full line marks the total 
solar abundance ratio ($\mathrm{[Ba/Eu]} \equiv 0$), and the dashed line marks the solar r-process 
fractional ratio $\mathrm{[Ba/Eu]}_r = -0.70$, according to \citet{Arlandini1999}. 
The ratio [Ba/Eu] can be used as a test of the relative importance of the s- and r-processes to the 
initial mix of materials at the birth of the stars. A first inspection of Figure~\ref{FigBaEu} shows 
a very low dispersion of the [Ba/Eu] ratio in contrast to that found by 
\citet{Sneden1997}. It is also interesting to note that there is no systematic difference 
between the ratios found for the stars in different evolutionary phases. We obtain a mean value\footnote{The mean values for the [Ba/Eu] and [Sr/Ba] ratios haven been computed using the individual [Ba/Eu] and [Sr/Ba] values for each sample star and are not the simple difference between the mean [Ba/Fe], [Eu/Fe] and [Sr/Fe] ratios.} 
of $\mathrm{[Ba/Eu]} = -0.18 \pm 0.09$ slightly higher than the values given by 
\citet{Sneden1997} for M~15. This ratio is still in the range --0.2 to --0.6 found for 
field halo stars and globular cluster (M~5, M~13, M~92) giants 
\citep{Gra1991,Gra1994,Armos1994,Shet1996}. The [Ba/Eu] ratios found in NGC~6752 are significantly 
higher than the solar pure r-process ratio. This shows that a part of the Ba abundance measured in 
these stars has also an s-process origin.

Strontium is mainly built during the double-shell burning phase of low and intermediate-mass AGB stars, 
but a significant fraction ($\sim$20\%) could be produced from the He-core burning in massive stars 
\citep{Rait1993,Pran1990}. On the other hand, barium comes mainly from AGB stars in the mass range 
$\sim$1--4 $M_{\odot}$ \citep{Busso1999}. 
It is therefore interesting to plot the ratio of these two elements 
as a function of the metallicity. In Figure~\ref{FigSrBa} we have plotted [Sr/Ba] ratios as a function 
of metallicity for our sample. 
As the weak s component is of secondary origin at low metallicity, the [Sr/Ba] ratios have to be compared 
to the solar pure r-process ratio. We have computed the solar r-process fraction $\mathrm{[Sr/Ba]}_r = -0.50$, 
which is displayed in Figure~\ref{FigSrBa}, following \citet{Arlandini1999}. A higher ratio is found if 
we include a weak s-process contribution to the solar Sr ($[\mathrm{Sr}_{r+w}/\mathrm{Ba}_r]=-0.10$). 
For our sample, we obtain a mean value of $\mathrm{[Sr/Ba]} = -0.12 \pm 0.21$ which 
is in good agreement with the ratios found in field halo stars in this metallicity range 
\citep[e.g.][]{Burris2000}. The subgiants have a lower mean value ($-0.25 \pm 0.07$) than the TO stars 
($0.01 \pm 0.22$) for the [Sr/Ba] ratio, but the scatter in the dwarfs is higher, so that the ratios found in 
both populations are compatible with an unique mean value. 
In any case, these ratios show clearly an s-process contribution to the synthesis of the n-capture 
elements in NGC~6752.

\subsection{Comparison with field stars}

\begin{figure}
  \centering
  \includegraphics[width=8cm]{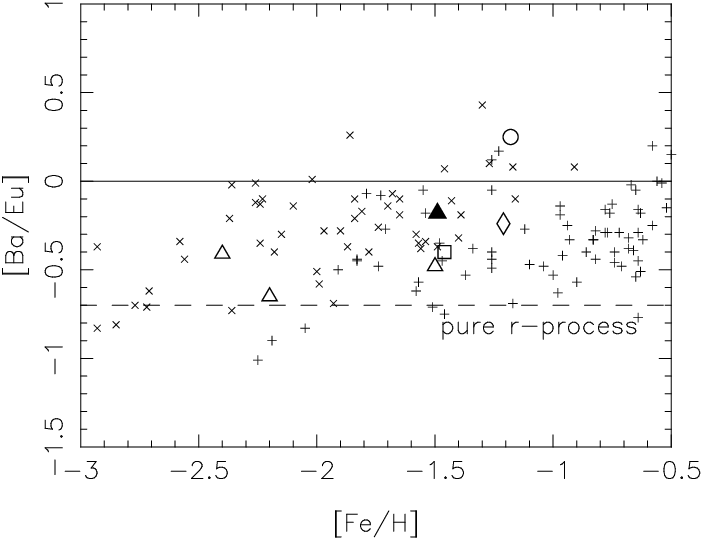}
  \caption{Comparison of [Ba/Eu] ratios for globular clusters and field halo stars. We show here the 
    mean value for our sample stars in NGC~6752 (black triangle). Other clusters are represented by 
    open symbols (with increasing metallicity: M~15, M~92, M~13, M~79, M~5 and M~4). 
    The abundances ratios for globular clusters 
    have been taken in \citet{Francois1991}, \citet{Sneden1997}, \citet{Ivans1999}, \citet{Ivans2001}, 
    and references therein. The field stars (crosses) are from \citet{Burris2000} and \citet{Fulb2000}.}
  \label{FigBaEuGCs}
\end{figure}

Several studies of barium and europium in Galactic field halo metal-poor stars 
\citep{McW1995,McW1998,Burris2000,Fulb2000} have shown that the [Ba/Eu] ratio falls with decreasing metallicity, 
and approaches a pure r-process ratio in most of the very metal-poor stars. Therefore it is interesting 
to compare what has been found in globular clusters stars and field halo stars in the same metallicity 
range. 

In Figure~\ref{FigBaEuGCs} we plot the mean [Ba/Eu] ratios for our cluster NGC~6752, and other globular 
clusters taken in the literature \citep[][ and references therein]{Francois1991,Sneden1997,Ivans1999}. 
We have added in this figure the data from \citet{Burris2000} and \citet{Fulb2000} for field stars. 
Our mean value $\mathrm{[Ba/Eu]} = -0.18 \pm 0.11$ for a metallicity of $\mathrm{[Fe/H]} = -1.49  \pm 0.07$ 
lies between the solar r+s mix and a pure r-process ratio, showing clearly an s-process contribution. Our 
mean values for [Ba/Eu] and [Sr/Ba] can also be compared to the much lower values found in the r-element 
rich stars CS~31082-001 ($\mathrm{[Ba/Eu]} = -0.46$, $\mathrm{[Sr/Ba]} = -0.52$, \citealt{Hill2002}) and 
CS~22892-052 ($\mathrm{[Ba/Eu]} = -0.65$, $\mathrm{[Sr/Ba]} = -0.39$, \citealt{Sneden2003}). 

Our data point for NGC~6752 is in good agreement with the field halo stars with the same metallicity. 
Although the dispersion is rather large, the [Ba/Eu] ratios found in globular clusters 
\citep{Francois1991,Sneden1997,Sneden2000,Ivans1999,Ivans2001} of different metallicities seem to follow 
the trend found in halo stars (i.e. [Ba/Eu] decreasing with decreasing metallicity). However, 
if we discard the high [Ba/Eu] value found in M~4 \citep{Ivans1999}, the data agrees also with a constant 
value of $\sim -0.4$. It is therefore very important to obtain new data for both the most metal-rich and 
metal-poor clusters.

\subsection{Self-pollution and self-enrichment of NGC~6752}

The recent results from \citet{Gra2001} have shown that the O-Na and Mg-Al anticorrelations found in giant 
stars of globular clusters are also present in main sequence stars of NGC~6752. These new data can be interpreted 
in the light of a scenario of self-pollution \citep{Cottrell1981}. In this scenario, the inhomogeneities are 
due to the mass lost by intermediate-mass stars during the AGB and planetary nebulae phases. 
\citet{Ventura2001} performed computations of self-pollution coming from low-metallicity AGBs, and showed 
that the observed O-Na and Mg-Al anticorrelations can be explained by the full CNO-cycle operating at the 
base of the envelope of the most metal-poor models of 4 and 5 $M_{\sun}$. In these models, 
the \element[][16]{O} is reduced and the sodium and aluminium production by proton-capture can occur. 
These models are only able to explain the anticorrelation of light metals, but not the origin of the 
metallicity of globular clusters, nor the heavy elements ratios. It is not yet clear whether globular 
clusters are formed out of matter which is already enriched. 

Different authors \citep{Cayrel1986,Truran1991,Parm1999} proposed models for the self-enrichment of globular 
clusters. In these models, a first generation of stars 
is assumed to form in the central region of the progenitor cloud. When the massive stars of this first 
generation exploses as SNeII, the matter of the globular cluster is enriched in iron, $\alpha$-elements 
and heavy elements. If the matter used to form the cluster is of primordial origin, the resulting [Ba/Eu] or 
[Sr/Ba] ratios should be compatible with pure r-process values. 

Our new data could be used as a test of these scenarii. However, there is no detailed computation 
concerning the self-enrichement of globular clusters. 
It is interesting to note that we found a [Ba/Eu] and [Sr/Ba] ratios departing significantly from the pure r-process 
ratios, excluding the self-enrichment scenario based only on the contribution of SNeII ejectae. Our values could be 
explained in the framework of the self-enrichement scenario, but only if an s-process contribution from low-mass 
AGB stars \citep{Busso1999} is included. 
However, the Ba abundances are constant among the observed stars in NGC~6752, in spite of the large 
star-to-star variation of the O and Na abundances, as found by \citet{Gra2001}. 
A similar result was found by \citet{Armos1994} for several other clusters, 
and later confirmed by other investigators. This indicates that the stars 
responsible for the O-Na anticorrelation did not produce significant amounts of s-process elements. 
It is also interesting to remind that we did not find any trend for any of the ratios [n-capture/Fe] as a 
function of [Na/Fe] or [O/Fe]. 

On the other hand, the fact that our abundance ratios are similar to the values found in halo stars 
with the same metallicity could be interpreted with a model where the globular cluster is formed out 
of matter which is already enriched in r- and s-process elements. In the framework of this last scenario, 
the n-capture elements abundances would have been present at the formation of the cluster, but the 
dispersion in O and Na could still be explained by massive AGBs.

\section{Conclusions}

In this paper, we presented the first abundance determinations for heavy elements in turn-off 
stars and early subgiants in NGC~6752. These results appear to be the first precise [Sr/Fe], [Y/Fe], 
[Ba/Fe], and [Eu/Fe] determinations in this cluster. We did not find any large systematic effect between 
the abundances found in turn-off and subgiant stars (this paper), and giant stars \citep{Norris1995a}. 

We obtain the following mean abundances in our sample (turn-off stars and subgiants): 
$$\mathrm{[Sr/Fe]} = 0.06 \pm 0.16$$
$$\mathrm{[Y/Fe]} = -0.01 \pm 0.12$$
$$\mathrm{[Ba/Fe]} = 0.18 \pm 0.11$$
$$\mathrm{[Eu/Fe]} = 0.41 \pm 0.09$$
Our results are in agreement with constant abundance ratios, and the low scatter can be totally 
explained by uncertainties in their derivation. These ratios are in agreement 
with the results found in field halo stars with the same metallicity. 

We did not observe any correlation between the [n-capture/Fe] ratios and the star-to-star variations 
of the O and Na abundances. 

Our mean values $\mathrm{[Ba/Eu]} = -0.18 \pm 0.11$ and $\mathrm{[Sr/Ba]} = -0.21 \pm 0.12$ lie both between 
a pure r-process and the solar r+s mix ratios. Looking at these ratios, we showed that NGC~6752 has been 
polluted by s-process nucleosynthesis. Whether this s-process signature comes from an internal enrichment or 
is due to a pre-enrichment of the matter from which the cluster formed, remains still an open question. 

It would be interesting to extend this kind of analysis to other globular clusters, 
especially the most metal-rich. Is is only with a significant set of data that the self-enrichement 
scenario can be tested, although no detailed modelisation has yet been done.

\begin{acknowledgements}
  The authors would like to thank the referee for his advices and his important contribution to the 
  discussion. G.J. would like to thank R. Cayrel, M. Spite, V. Hill, and E. Depagne (Observatoire de 
  Paris, France), and S. Lefranc (IAS, France) for the many helpful discussions. 
  E.C., S.D., R.G.G., and S.L. research was funded by COFIN 2001028897 by Ministero 
  Universita' e Ricerca Scientifica, Italy. 
\end{acknowledgements}

\bibliographystyle{aa}


\end{document}